\documentclass[prb,twocolumn,showpacs]{revtex4}
\usepackage{graphicx}

\begin{document}

\title{Strain effect on electronic transport and ferromagnetic transition temperature 
in La$_{0.9}$Sr$_{0.1}$MnO$_{3}$ thin films }
\author{X. J. Chen}
\affiliation{Max-Planck-Institut f\"{u}r Festk\"{o}rperforschung, D-70569 Stuttgart, Germany\\
and Department of Physics, Kent State University, Kent, OH 44242, USA } 
\author{S. Soltan, H. Zhang, and H.-U. Habermeier}
\affiliation{Max-Planck-Institut f\"{u}r Festk\"{o}rperforschung, D-70569 Stuttgart, Germany} 
\date{Received 13 December 2001}

\begin{abstract}
We report on a systematic study of strain effects on the transport properties and the 
ferromagnetic transition temperature $T_{c}$ of high-quality La$_{0.9}$Sr$_{0.1}$MnO$_{3}$ 
thin films epitaxially grown on (100) SrTiO$_{3}$ substrates. Both the magnetization and 
the resistivity are critically dependent on the film thickness. $T_{c}$ is enhanced with 
decreasing the film thickness due to the compressive stain produced by lattice mismatch. 
The resistivity above 165 K of the films with various thicknesses is consistent with small 
polaronic hopping conductivity. The polaronic formation energy $E_{P}$ is reduced with 
the decrease of film thickness. We found that the strain dependence of $T_{c}$ mainly results 
from the strain-induced electron-phonon coupling. The strain effect on $E_{P}$ is in good 
agreement with the theoretical predictions.
\end{abstract}

\maketitle

\section{Introduction}

The discovery of colossal magnetoresistance effect in epitaxial manganite thin films has 
renewed interest in the doped manganite perovskite materials Ln$_{1-x}$B$_{x}$MnO$_{3}$ 
(Ln=trivalent rare-earth ions; B=divalent alkaline-earth ions) for potential sensor and 
magnetic recording applications as well as the need to understand the mechanisms underlying 
their behavior.\cite{helm,sjin,hlju} It has been found that properties such as ferromagnetic 
transition temperature $T_{c}$, resistivity $\rho$, and magnetoresistance are sensitive to 
the epitaxial strain due to lattice mismatch of the film with substrate.
\cite{sjin2,prel,koo,kwon,raza,arao,shre} 
When the film is grown on the substrate whose lattice parameter is smaller or larger than 
that of the bulk material, the epitaxial strain is expected to be compressive or tensile, 
respectively. Compressive strain usually reduces the resistivity and shifts $T_{c}$ towards 
the higher temperature. These effects have been confirmed in La$_{0.7}$Ca$_{0.3}$MnO$_{3}$ 
films\cite{koo} and La$_{0.7}$Sr$_{0.3}$MnO$_{3}$ films\cite{kwon} grown on various substrates.

The observed strain effect is usually interpreted qualitatively within double exchange model,
\cite{genn} since the hopping matrix element $t$ could be altered by epitaxial strain 
through changing the Mn-O bond length $d$ and the Mn-O-Mn bond angle $\theta$. It has 
been also proposed that the Jahn-Teller electron-phonon coupling plays an important 
role in strain effect on $T_{c}$.\cite{mill} However, recent detailed studies show that  
compressive strain does not always lead to enhancement of $T_{c}$,\cite{arao} while the 
cationic vacancies due to the oxygen annealing significantly enhance the $T_{c}$ values much 
higher than any bulk values in the series compounds.\cite{prel,shre} In most cases, tensile 
strain suppresses ferromagnetism and reduces $T_{c}$ in manganite films. But some anomalous 
results have also been reported, showing $T_{c}$ enhanced by tensile strain.
\cite{gong,prel2,zhan} Most interestingly, there are reports of multiple phase segregation 
in fully strained epitaxial films.\cite{bibe} The ferromagnetic coupling within the metallic 
regions accounts for the changes of $T_{c}$ and conductivity. Thus, the strain effect in manganite 
films is far from being fully understood and challenging.

Lightly doped La$_{1-x}$Sr$_{x}$MnO$_{3}$ shows a great variety of intriguing phenomena 
originating from a pronounced interplay between spin, lattice, charge, and orbital degrees 
of freedom. As a result many phenomena like charge order,\cite{yama,uhle} orbital order,
\cite{endo} and phase separation\cite{zhou} have been recently observed in this regime of 
the phase diagram. La$_{0.9}$Sr$_{0.1}$MnO$_{3}$ is in the phase boundary of a spin-canted 
antiferromagnetic insulator and a ferromagnetic insulator.\cite{woll,urus,kawa,dabr} 
This material has the lowest $T_{c}$ among the series compounds,\cite{urus,morit} which 
makes it possible to perform systematic investigations of the resistivity in the paramagnetic 
regime over a broad temperature range without using specialized equipment to extend the 
temperature range. Meanwhile, the pressure derivative of $T_{c}$, $dT_{c}/dP$, in this 
material is highest among the manganese perovskites.\cite{morit,senis,tiss} It has been 
generally believed that pressure changes $T_{c}$ and $\rho$ in a similar manner as epitaxial 
strain. Thus, transport properties, transition temperatures, and phase transitions are 
expected to be significantly affected by epitaxial strain in the La$_{0.9}$Sr$_{0.1}$MnO$_{3}$ 
films. Moreover, these investigations are most important for the understanding of the fruitful 
phenomena and the use of these films as magnetic devices as well as air electrodes in 
high-temperature solid oxide fuel cells.\cite{roos,wolf0}

In this work we investigate the transport properties by measuring resistivity and magnetization 
of the epitaxial La$_{0.9}$Sr$_{0.1}$MnO$_{3}$ films on SrTiO$_{3}$. The data clearly show 
that the high-temperature resistivity of the films can be well ascribed by a model for 
small-polaron hopping in the adiabatic limit. We experimentally find that the small polaronic 
formation energy $E_{P}$ decreases with the reduction of the film thickness, which can 
account for the strain effect on $T_{c}$. We suggest that the electron-phonon coupling is 
responsible for the strain effect on the high-temperature electronic transport and the 
ferromagnetic transition temperature.

\section{Experimental details}

Thin films of La$_{0.9}$Sr$_{0.1}$MnO$_{3}$ were grown using the pulsed laser deposition technique. 
The target used had a nominal composition of La$_{0.9}$Sr$_{0.1}$MnO$_{3}$. The substrates were 
(100) single crystal of SrTiO$_{3}$. The laser energy density on the target was 2 mJ/cm$^{2}$ and 
the ablation rate was 5 Hz. The substrates were kept at a constant temperature of 850$^{0}$C during 
the deposition which was carried out at a pressure of 0.40 mbar of oxygen. The films were $in$ $situ$ 
annealed at 940$^{0}$C in oxygen at 1.0 bar for 30 minutes. This procedure always results in films of 
high crystalline quality and in very sharp film-substrate interfaces. The thickness of the films was 
varied from 200 to 2000 $\AA$ as measured by Dektak. The chemical composition of the films was 
determined by microprobe analysis, which showed a (La,Sr)/Mn ratio of 1:1 and a Sr content of 
$x=0.10\pm 0.01$.

The structural study was carried out by x-ray diffraction (XRD) at room temperature by a 
Rigaku x-ray diffractometer with a rotating anode and Cu $K\alpha$ radiation, 
$\lambda=1.5406$ $\AA$. The resistivity $\rho$ was measured from unpatterned samples with 
sputtered chromium gold contacts using a standard four-probe technique. Magnetization $M$ was 
recorded in a magnetic field parallel to the film plane using a Quantum Design MPMS superconducting 
quantum interference device (SQUID) magnetometer as a function of temperature.

\section{Results and discussion}

\begin{figure}[t]
\begin{center}
\includegraphics[scale=0.32]{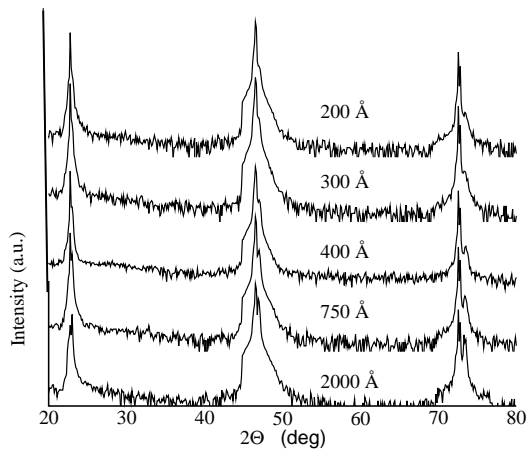}
\end{center}
\caption{ Room temperature XRD of La$_{0.9}$Sr$_{0.1}$MnO$_{3}$ films for various thicknesses. }
\label{fig1}
\end{figure}

Figure \ref{fig1} shows the evolution of the room temperature XRD data for La$_{0.9}$Sr$_{0.1}$MnO$_{3}$ 
thin films with thicknesses from 200 to 2000 $\AA$. Each sample is single crystal and ($l00$) 
oriented without other impurity phases. Above $T=105$ K, SrTiO$_{3}$ has a perfect cubic 
perovskite structure with a lattice parameter $a=3.905$ $\AA$. La$_{0.9}$Sr$_{0.1}$MnO$_{3}$ 
has a distorted perovskite structure due to the tilting of the MnO$_{6}$ octahedra and the 
Jahn-Teller distortion, which results in a slightly orthorhombic structure. The bulk lattice 
parameters for this compound at room temperature are:\cite{cox} $a=5.5469$ $\AA$, $b=5.56033$ 
$\AA$, and $c=7.7362$ $\AA$. The in-plane lattice mismatch between the film and the substrate 
is given by $\epsilon=[d_{bulk}-d_{strained}]/d_{bulk}$ with $d$ a lattice parameter. 
Epitaxially grown La$_{0.9}$Sr$_{0.1}$MnO$_{3}$ film on (100) SrTiO$_{3}$ substrates are under 
compressive strain since $d_{bulk}>d_{strained}$ with the bulk value $d_{bulk}=3.927 \AA$. With 
decreasing the film thickness, the in-plane lattice parameter of the film decreases and the 
compressive strain is then enhanced.

\begin{figure}[t]
\begin{center}
\includegraphics[scale=0.32]{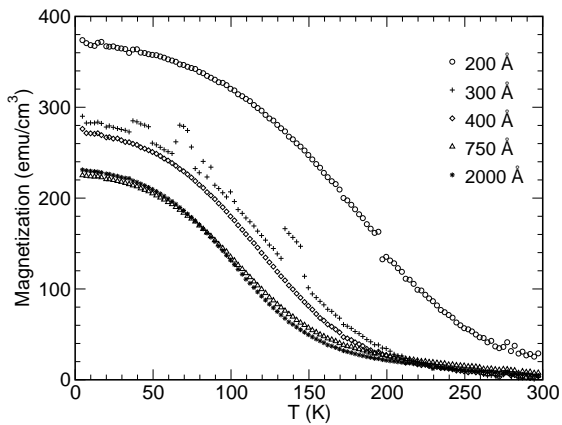}
\end{center}
\caption{Magnetization as a function of temperature measured in a field of 0.5 T of 
La$_{0.9}$Sr$_{0.1}$MnO$_{3}$ films with various thicknesses. }
\label{fig2}
\end{figure}

Figure \ref{fig2} shows the temperature dependence of the magnetization measured in 0.5 T of 
the films with various thicknesses, after correction for the magnetization of the substrate.
The curves have been measured by warming up in the magnetic field after zero field cooling. 
The features of the $M-T$ curves are ferromagnetic with $M\sim 230-360$ emu/cm$^{3}$ at 10 K.
The magnetization was taken at 0.5 T to avoid the variation due to magnetic domain rotation 
at lower fields. Both $T_{c}$ and $M$ increase with decreasing the film thickness. The value 
of $T_{c}$ for 200 $\AA$ thin film is 50 K higher than the bulk value.\cite{senis} We had not 
observed a magnetization jump occurring at a characteristic temperature $T_{CA}$ as appeared 
in the La$_{0.9}$Sr$_{0.1}$MnO$_{3}$ single crystals,\cite{senis,koro} which indicates a 
canted antiferromagnetic state as confirmed by neutron scattering experiments.\cite{yama}        
This is not surprising since the strained films usually show properties much different from 
the bulk compounds in manganites.\cite{prel}

Although the reduction of film thickness should enhance $T_{c}$ under compressive strain as we 
observed in Fig. \ref{fig2}, there are few measurements in other manganites films to support this 
phenomenon. The experiments on La$_{0.8}$Ca$_{0.2}$MnO$_{3}$ films grown on LaAlO$_{3}$ do not 
always show a correlation between the compressive strain and $T_{c}$.\cite{arao} Interestingly, 
anomalously high $T_{c}$ and metal-insulator transition temperature $T_{MI}$ (100 K higher than 
bulk values) have been observed in this strained film with 1000 $\AA$ thickness after annealing 
under oxygen.\cite{shre} For this La$_{0.8}$Ca$_{0.2}$MnO$_{3}$ film, $T_{MI}$ is 30 K higher 
than the highest $T_{MI}=260$ K found for $x=0.33$ bulk compound. Thus, the large enhancement 
of $T_{c}$ and $T_{MI}$ in this film should be dominated by compressive strain. The lack of this 
enhancement observed previously in La$_{0.8}$Ca$_{0.2}$MnO$_{3}$ thin film may be due to oxygen 
deficiency.

\begin{figure}[b]
\begin{center}
\includegraphics[width=\columnwidth]{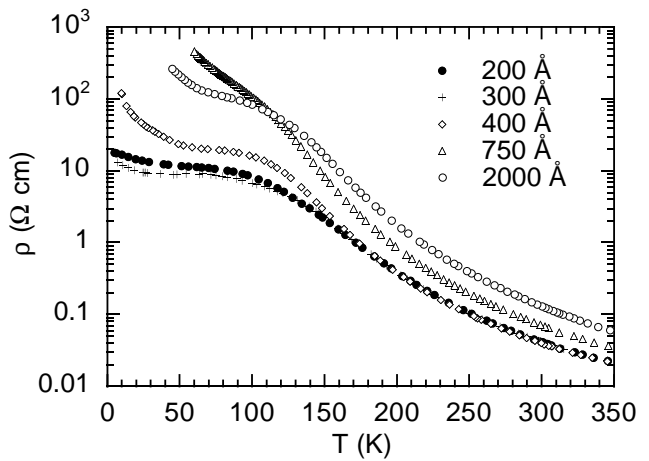}
\end{center}
\caption{Temperature dependence of resistivity of La$_{0.9}$Sr$_{0.1}$MnO$_{3}$ films with various 
thicknesses. }
\label{fig3}
\end{figure}

The results of the temperature dependence of the resistivity are shown in Fig. \ref{fig3}. The 
resistivity of our films displays semiconducting behavior at high temperatures, and metallic 
behavior for $T_{CA}\leq T \leq T_{MI}$. It has an upturn at $T_{CA}$, and then 
becomes of semiconducting character. The neutron scattering study demonstrates that the 
point of resistivity upturn is consistent with the onset temperature of the polaron order.
\cite{yama} The magnitude of resistivity of our films is smaller than those of the single 
crystals.\cite{urus,dabr,liu} For example, the resistivity of the 2000 $\AA$ film at $T=100$ 
K is 83.7 $\Omega $ cm. Note that the compressive strain decreases the resistivity in our 
thin films. This behavior is typical for manganites films under compressive strain.\cite{koo,kwon} 
The observed $T_{MI}$ (defined as the temperature where $d\rho/dT$ changes sign) of 
$\sim 100-150$ K in our films are comparable to those of La$_{0.9}$Sr$_{0.1}$MnO$_{3}$ 
single crystals.\cite{urus,dabr,liu} For films with thicknesses $d=750$ and 2000 $\AA$, 
$T_{MI}$ almost coincides with $T_{c}$. However, $T_{MI}$ is significant smaller than $T_{c}$ 
for the ultrathin films. The scenario to correlate with this observation could be the existence of 
microscopic phase segregation due to the formation of small ferromagnetic clusters, which are 
large enough to give a magnetic contribution in ultrathin films but not to allow metallic 
conductivity appearing in zones of ferromagnetic insulating behavior. The smaller $T_{MI}$ 
value compared to $T_{c}$ has reported previously in La$_{0.67}$Sr$_{0.33}$MnO$_{3}$ thin films.
\cite{gonz} Recent nuclear magnetic resonance measurements in La$_{2/3}$Ca$_{1/3}$MnO$_{3}$ 
films on SrTiO$_{3}$ give strong evidence in favor of the existence of microscopic phase 
separation.\cite{bibe}

Additional increase of resistivity on cooling can be seen at low temperatures proceeded 
by a minimum at $T_{CA}$. The structural data of single crystals show that a phase 
transformation from a pseudocubic $O^{\prime\prime}$ type to an orthorhombic $O^{\prime}$ 
type structure occurs near $T_{CA}$.\cite{kawa,gavi}The low-temperature phase is known to be 
a spin-canted antiferromagnetic phase for $0\leq x\leq 0.1$,\cite{woll} which results from 
competing antiferromagnetic superexchange interaction between half-filled $t_{2g}$ orbitals 
along the $c$ axis Mn-O-Mn bonds and ferromagnetic double-exchange interaction via $e_{g}$ 
conduction electrons. With the reduction of the film thickness, $T_{CA}$ shifts towards low 
temperatures and $\rho$ decreases in the insulating low-temperature phase. It has been 
reported that $T_{CA}$ increases and $\rho$ decreases under pressure in 
La$_{0.9}$Sr$_{0.1}$MnO$_{3}$ single crystals.\cite{senis} Although $\rho$ behaves in a similar 
manner under compressive strain and external pressure, the observed variation of $T_{CA}$ is 
in sharp contrast with the pressure measurements. It has been established\cite{zhou} that 
pressure influences $T_{CA}$ in the same way as an increase in $x$ with a maximum within the 
range $0.12<x<0.15$ for the slightly doped La$_{1-x}$Sr$_{x}$MnO$_{3}$. At low-pressures, the 
thermoelectric power through $T_{CA}$ is sensitive to the charge carrier density. It is indicated 
that pressure induces the change of carrier concentration, which should account for the 
dependence of $T_{CA}$ on pressure. The growth conditions such as film deposition and oxygen 
annealing are same for all films studied here. The carrier concentration in these films should 
not be different. Therefore, the dependence of $T_{CA}$ on strain is possibly different from the 
pressure effect on $T_{CA}$.

An interesting feature is the absence of the jump in resistivity in films near $T\sim330$ K. 
The structural analyses on La$_{0.9}$Sr$_{0.1}$MnO$_{3}$ crystals\cite{dabr,cox,gavi} reveal 
that the system undergoes another structural transition around characteristic temperature 
$T_{s}=330$ K from an orthorhombic $O$ phase having a dynamic Jahn-Teller distortion to a 
orthorhombic $O^{\prime}$ phase at lower temperatures where Jahn-Teller distortion becomes 
static and cooperative. The jump in resistivity at $T_{s}$ in single crystals has been 
reported by Urushibara $et$ $al.$\cite{urus} The absence of the jump indicates that the 
compressive strain in films either suppresses the structural phase transition or shifts $T_{s}$ 
towards higher temperatures above 350 K. There is competition between the charge mobility 
and the structural phase transition in the slightly doped La$_{1-x}$Sr$_{x}$MnO$_{3}$.\cite{uhle} 
The change tendency of $T_{s}$ and $T_{CA}$ is usually different under pressure or 
magnetic field.\cite{uhle,zhou} In our films, $T_{CA}$ decreases with decreasing film 
thickness due to the compressive strain.  Thus, the increase of $T_{s}$ is possible under 
compressive strain.

\begin{figure}[t]
\begin{center}
\includegraphics[width=\columnwidth]{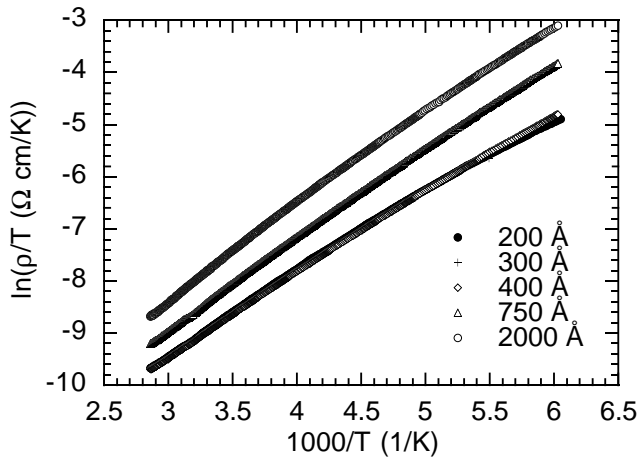}
\end{center}
\caption{Plot of $\ln (\rho/T)$ versus $1000/T$ of La$_{0.9}$Sr$_{0.1}$MnO$_{3}$ films with 
various thicknesses. }
\label{fig4}
\end{figure}

The preconditions for polaron formation, namely, large electron-lattice coupling and low 
electronic hopping rates, appear to be satisfied for manganites.\cite{millis} In Fig. \ref{fig4} 
we have represented $\ln(\rho/T)$ versus inverse temperature. A linear behavior is obtained 
between 165 and 350 K. This is strong support of the mechanism of adiabatic small polaron 
conduction. The resistivity as a result of hopping of adiabatic small polarons is, within 
the Emin and Holstein theory,\cite{emin} given by
\begin{equation}
\label{small}
\rho=AT\exp\left(\frac{E_{A}}{k_{B}T}\right)~~.
\end{equation}
Here the prefactor $A$ depends on the polaronic concentration, the hopping distance, and the
frequency of the longitudinal optical phonon. The activation energy $E_{A}$ has the form
\cite{jaime} $E_{A}=E_{P}/2+\epsilon_{0}-J$, where $\epsilon_{0}$ is the energy required to
generate intrinsic carriers and $J$ is the transfer integral.

From the fit to Eq. (\ref{small}), the values of $A$ and $E_{A}$ are obtained. These data are 
summarized in Table \ref{table1}. The fitting for $\rho$ is valid for temperatures larger than 
half the Debye temperature $\Theta_{D}$. This is fulfilled for the present films since specific 
heat measurements show $\Theta_{D}$ in the $255-360$ K range.\cite{wood,okud} We noted that the
fitting adiabatic prefactor $A$ is in the range from 1.19$\times 10^{-6}$ to 2.39$\times 
10^{-6}$ $\Omega$ cm/K, which is typical for small polaronic conduction as observed in 
La$_{0.67}$Ca$_{0.33}$MnO$_{3}$ films\cite{worl} as well as 
(La$_{1-x}$Gd$_{x}$)$_{0.67}$Ca$_{0.33}$MnO$_{3}$ films.\cite{jaime2}  

There have been studies of high-temperature resistive behavior in La$_{0.9}$Sr$_{0.1}$MnO$_{3}$ 
bulk materials.\cite{raff,para,mand} The reported conduction mechanism are controversial. 
Early measurements on the ceramic La$_{0.9}$Sr$_{0.1}$MnO$_{3}$ show that the high-temperature
resistivity obeys the small polaron transport behavior in the nonadiabatic limit with an 
activation energy $E_{A}=0.2$ eV.\cite{raff} In single crystals, some groups found that their 
data can be well fitted by variable-range-hopping model 
$\rho =\rho _{0}(T/T_{0})^{1/2}exp[(T_{0}/T)^{1/4}]$ with $T_{0}=1.72\times 10^{8}$ K in the 
paramagnetic regime,\cite{para} while others\cite{mand} reported the resistivity follows a 
small polaron model in adiabatic limit above $T_{MI}$ with activation energy $E_{P}=0.3$ eV. 
The high-temperature resistivity of our films with various thicknesses is consistent with 
adiabatic small polaron hopping conductivity. It has been generally accepted that the 
conductivity can be  well ascribed by adiabatic small polaron model in 
La$_{1-x}$Ca$_{x}$MnO$_{3}$ films.\cite{jaime,worl,jaime2,ziese,worl2,tere} Our present data provide 
clear support for the existence of this conductivity mechanism in La$_{1-x}$Sr$_{x}$MnO$_{3}$ 
films.  

\begin{table}[b]
\caption{ Thickness dependence of the activation energy $E_{A}$, the resistivity 
coefficient $A$, and the ferromagnetic transition temperature $T_{c}$ in 
La$_{0.9}$Sr$_{0.1}$MnO$_{3}$ films. }
\label{table1}
\begin{ruledtabular}
\begin{tabular}{cccc}
Thickness ($\AA$) & $E_{A}$ (meV) & A ($10^{-6}\Omega$ cm/K) & $T_{c}$ (K)\\
\hline
200 & 119.1 & 1.86 & 194.9 \\
300 & 124.8 & 1.28 & 150.0 \\
400 & 126.8 & 1.19 & 116.9 \\
750 & 139.6 & 1.28 & 100.0 \\
2000 & 141.3 & 2.39 & 105.6 \\
\end{tabular}
\end{ruledtabular}
\end{table}

\begin{figure}[t]
\begin{center}
\includegraphics[scale=0.35]{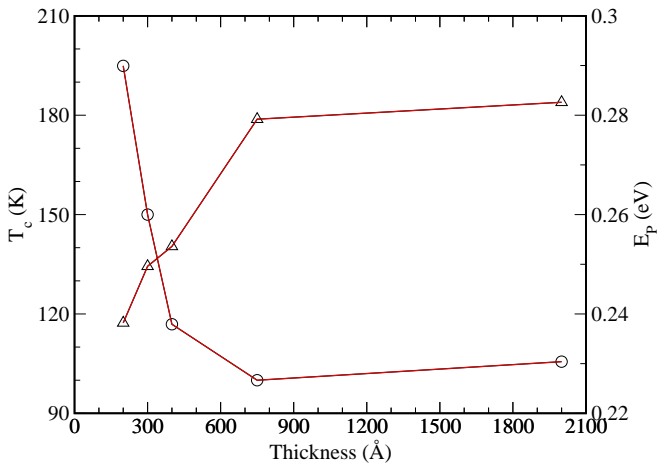}
\end{center}
\caption{Thickness dependence of the ferromagnetic transition temperature $T_{c}$ (circles) and the 
polaronic formation energy $E_{P}$ (triangles) in La$_{0.9}$Sr$_{0.1}$MnO$_{3}$ films. }
\label{fig5}
\end{figure}

At high temperatures and in the adiabatic limit the contribution from $\epsilon_{0}$ and
$J$ may be neglected, the variation of $E_{P}$ is approximately affected by the change of 
$E_{A}$. Taking $E_{P}=2E_{A}$, we have plotted the thickness dependence of $E_{P}$ in Fig. \ref{fig5}. 
The thickness dependence of $T_{c}$ is also plotted for comparison. It is interesting to 
notice that the variation of $T_{c}$ with thickness can be well reflected by the thickness 
dependence of $E_{P}$. For the thick films, the strain is relaxed. Both $T_{c}$ and $E_{P}$ 
scarcely change with the thickness. Below 750 $\AA$, with the reduction of film thickness, 
$E_{P}$ decreases, whereas $T_{c}$ increases. It is therefore indicated that the electron-phonon 
coupling possibly dominates the strain effect on $T_{c}$.

The polaronic formation energy $E_{P}$ is usually related to the effective bandwidth $W_{eff}$ 
in polaronic models. Zhao $et$ $al.$\cite{zhao} proposed an effective bandwidth of the 
form $W_{eff}=W\exp(-\gamma E_{P}/\hbar \nu)$, where $W$ is the electronic ``bare'' 
bandwidth, $\nu$ is the characteristic vibration frequency of the optical phonon mode, 
and $\gamma$ depends on the ratio $E_{P}/W$. According to the model proposed by Varma, 
\cite{varma} $T_{c}$ can be written as
\begin{equation}
\label{tc}
T_{c}=\frac{0.1}{2}W\exp\left(-\frac{\gamma E_{P}}{\hbar \nu}\right)n\left(1-n\right)~~,
\end{equation} 
where $n$ denotes the carrier concentration. Considering that $\nu$ is related to the isotope 
mass $M$ through $\nu \propto M^{-1/2}$, the oxygen isotope exponent $\alpha$ ($\equiv -d\ln 
T_{c}/d\ln M$) is then given by $\alpha=0.5\gamma E_{P}/\hbar \nu$. The strain coefficient of 
$T_{c}$, $d\ln T_{c}/d\epsilon$, is readily obtained from Eq. (\ref{tc})
\begin{equation}
\label{dtce}
\frac{d\ln T_{c}}{d\epsilon}=\frac{d\ln W}{d\epsilon}-2\frac{d\alpha}{d\epsilon}~~.
\end{equation} 

For La$_{0.9}$Sr$_{0.1}$MnO$_{3}$, the pressure coefficient of $T_{c}$ has been found by 
Senis $et$ $al.$\cite{senis} to be $d\ln T_{c}/dP=0.16$ GPa$^{-1}$. Using the lattice 
compressibility $\kappa_{a}=2.32\times 10^{-3}$ GPa$^{-1}$,\cite{rada} we obtain $d\ln 
T_{c}/d\epsilon=69$. The electronic bandwidth $W$ of the manganites can be estimated by the 
average Mn--O bond distance $d$ and the Mn-O-Mn angle $\theta$ by using the relation:\cite{meda} 
$W \propto \cos \phi/d^{3.5}$, where $\phi=(\pi-<\theta>)/2$. The pressure dependence of 
$\cos \phi$ has been determined by neutron diffraction measurements\cite{rada} to be 
$(\cos\phi)^{-1}d\cos\phi/dP=2.1\times 10^{-4}$ GPa$^{-1}$. Taking the value of $\kappa_{a}$ 
as the bond compressibility $\kappa_{d}$, the calculated $d\ln W/d\epsilon$ is 3.6. Thus, 
$d\alpha/d\epsilon=-32.7$ is obtained from Eq. (\ref{dtce}). In La$_{0.9}$Sr$_{0.1}$MnO$_{3}$, the 
oxygen isotope exponent $\alpha=0.2$ reported previously by Zhao $et$ $al.$\cite{zhao} Based 
on the above determined parameters, one estimated the pressure derivate of $\alpha$, 
$d\alpha/dP=-0.076$ GPa$^{-1}$. This value is very close to the reported value of --0.05 
GPa$^{-1}$ in La$_{0.65}$Ca$_{0.35}$MnO$_{3}$.\cite{wang}

According to the expression for $\alpha$, $d\alpha/d\epsilon$ is then expressed as
\begin{equation}
\label{dae}
\frac{d\alpha}{d\epsilon}=\alpha\left(\frac{d\ln E_{P}}{d\epsilon}-\frac{d\ln 
\nu}{d\epsilon}\right)~~.
\end{equation} 
The Raman spectra of La$_{0.9}$Sr$_{0.1}$MnO$_{3}$ have been collected previously by 
Podobedov $et$ $al.$\cite{podo} The sharp peaks at the top of the wide band are located at 
243, 493, and 609 cm$^{-1}$. The high frequency $B_{1g}$ mode at 609 cm$^{-1}$ is suggested 
as a stretching Mn-O vibration. Recent high pressure studies\cite{cong} show that this 
stretching mode is the most sensitive to pressure with an initial pressure coefficient, 
$d\ln \nu/dP=0.01$ GPa$^{-1}$. Thus $d\ln \nu/d\epsilon=4.4$. Equation (\ref{dae}) gives
$d\ln E_{P}/d\epsilon=-159$. This follows that $E_{P}$ decreases with increasing 
compressive strain. This is in good agreement with our experimental fitting parameters 
as shown in Fig. 5. Combining Eqs. (\ref{dtce}) and (\ref{dae}), we can conclude that the 
strain dependence of $T_{c}$ mainly results from the strain dependence of the polaronic 
formation energy though there are also contributions from the electronic bandwidth $W$ and 
the characteristic phonon frequency $\nu$.

\section{Conclusions}

We have epitaxially grown La$_{0.9}$Sr$_{0.1}$MnO$_{3}$ thin films on SrTiO$_{3}$ substrates. 
The high-temperature resistivity of the films with various thicknesses obeys the small-polaron 
hopping conductivity in the adiabatic limit. We experimentally find that the small polaronic 
formation energy $E_{P}$ decreases with the reduction of the film thickness, which mainly
accounts for the the strain effect on $T_{c}$. By theoretical analysis, we found the 
contribution from electronic bandwidth is much smaller than that from electron-phonon
interaction.  We therefore concluded that the electron-phonon coupling is responsible for 
the strain effect on the high-temperature electronic transport and the ferromagnetic transition 
temperature in our films.

\end{document}